# Finite-SNR Diversity-Multiplexing Tradeoff via Asymptotic Analysis of Large MIMO Systems

Sergey Loyka, *Senior Member, IEEE*, Georgy Levin

*Abstract*—Diversity–multiplexing tradeoff (DMT) was characterized asymptotically (SNR-> infinity) for i.i.d. Rayleigh fading channel by Zheng and Tse [1]. The SNR-asymptotic DMT overestimates the finite-SNR one [2]. This paper outlines a number of additional limitations and difficulties of the DMT framework and discusses their implications. Using the recent results on the size-asymptotic (in the number of antennas) outage capacity distribution, the finite-SNR, size-asymptotic DMT is derived for a broad class of fading distributions. The SNR range over which the finite-SNR DMT is accurately approximated by the SNR-asymptotic one is characterized. The multiplexing gain definition is shown to affect critically this range and thus should be carefully selected, so that the SNR-asymptotic DMT is an accurate approximation at realistic SNR values and thus has operational significance to be used as a design criteria. The finite-SNR diversity gain is shown to decrease with correlation and power imbalance in a broad class of fading channels, and such an effect is described in a compact, closed form. Complete characterization of the outage probability (or outage capacity) requires not only the finite-SNR DMT, but also the SNR offset, which is introduced and investigated as well. This offset, which is not accounted for in the DMT framework, is shown to have a significant impact on the outage probability for a broad class of fading channels, especially when the multiplexing gain is small. The analytical results and conclusions are validated via extensive Monte-Carlo simulations. Overall, the size-asymptotic DMT represents a valuable alternative to the SNR-asymptotic one.

## I. Introduction

MULTI-antenna (MIMO) systems are able to provide either high spectral efficiency (spatial multiplexing) or low error rate (high diversity) via exploiting multiple degrees of freedom available in the channel, but not both simultaneously as there is a fundamental tradeoff between the two. This diversity-multiplexing tradeoff (DMT) is best characterized using the concepts of multiplexing and diversity gains [1]. Fundamentally, this is a tradeoff between the outage probability $P_{out}$, i.e. the probability that the fading channel is not able to support the transmission rate $R$, and the rate $R$, which can be expressed via the outage capacity distribution,

$$P_{out}(R) = \Pr[C < R] = F_C(R) \quad (1)$$

where $C$ is the instantaneous channel capacity (i.e. capacity of a given channel realization), and $F_C(R)$ is its cumulative distribution function (CDF). Defining the multiplexing gain $r$ as

$$r = \lim_{\gamma \to \infty} R/\ln\gamma \quad (2)$$

where $\gamma$ is the average SNR at the receiver, and the diversity gain as[1]

$$d = -\lim_{\gamma \to \infty} \frac{\ln P_{out}}{\ln \gamma} \quad (3)$$

the SNR-asymptotic ($\gamma \to \infty$) tradeoff for the independent identically distributed (i.i.d.) Rayleigh fading channel with the coherence time in symbols $T \geq m + n - 1$ can be compactly expressed as [1],

$$d(r) = (n-r)(m-r), \ r = 0, 1, ... \min(m,n) \quad (4)$$

where $m, n$ are the number of transmit (Tx), receive (Rx) antennas, for integer values of $r$, and using the linear interpolation in-between. The motivation for the definition of $r$ in (2) is that the mean (ergodic) capacity $\overline{C}$ scales as $\min(m,n)\ln\gamma$ at high SNR,

$$\overline{C} \approx \min(m,n)\ln\gamma, \text{ as } \gamma \to \infty \quad (5)$$

and the motivation for the definition of $d$ in (3) is that $P_{out}$ scales as $\gamma^{-d}$ at high SNR,

$$P_{out} \approx c/\gamma^d, \text{ as } \gamma \to \infty \quad (6)$$

where $c$ is a constant independent of the SNR[2]. The DMT in (4) has been extended to multiple-access channels in [3].

While the SNR-asymptotic approach provides a significant insight into MIMO channels and also into performance of various systems that exploit such channels, it has also a number of limitations. Specifically, it does not say anything about operational significance of $r$ and $d$ at realistic (i.e. low to moderate) SNR. In other words, how high is the SNR required to approach the asymptotes in (2), (3), with reasonable accuracy, so that, for example, $d$ can be used to accurately estimate $P_{out}$ using (6) and (4)? It was observed in [2], based on a lower bound on $P_{out}$ for Rayleigh and Rician channels, that the finite-SNR DMT lies well below the curve in (4), and the convergence of the finite-SNR DMT to the asymptotic value in (4) as the SNR grows is slow, so that (4) becomes an accurate approximation for the finite-SNR DMT only at unrealistically high SNR. Therefore, proper modifications to the asymptotic results and definitions are required to improve its accuracy for realistic SNR values. Using the SNR-asymptotic ($\gamma \to \infty$) DMT to compare two systems may give incorrect results at low to moderate SNR. The estimates of finite-SNR DMT based on lower bounds on the outage probability in correlated Rayleigh and Rician channels obtained in [2] are not in an explicit closed-form (e.g. (4)) and require a numerical procedure to evaluate, which limits the insight that can be extracted from such results.

---

[1] while the original definition in [1] employed the average error rate, the definition in (3) is equivalent to it since the average error rate is dominated by the outage probability [35]. This definition has also been used in [2].

[2] but, as we demonstrate later on, this constant is a function of the multiplexing gain and limits the applicability of the SNR-asymptotic DMT, even at high SNR, for some systems.

Another approach to the problem has been presented in [4], [5], where the rate is not required anymore to satisfy the condition in (2), but rather belongs to a rate region, $k < R/\ln\gamma < k+1$, $k = 0, 1, \ldots \min(m, n)$. Based on the concept of rate regions, it has been demonstrated that there exists a tradeoff between the outage probability and the rate termed "throughput-reliability tradeoff (TRT)", which can be expressed, for the i.i.d. Rayleigh fading channel, in a compact form as

$$\lim_{\gamma \to \infty} \frac{\ln P_{out} - c(k)R}{\ln \gamma} = -g(k), \quad (7)$$

where $g(k) = mn - k(k+1)$, $c(k) = m + n - (2k+1)$. While the rigorous result still requires $\gamma \to \infty$, the TRT is more accurate at finite SNR values compared to (4), and it provides a reasonably-accurate finite-SNR answer to the important question "what does a 3dB buy in MIMO channels?" [5]. However, (7) does overestimate the outage probability at low to moderate SNR values [4].

While the original DMT formulation of Zheng and Tse is limited to i.i.d. Rayleigh fading channels, a generalization to a class of channels satisfying a number of conditions on the distribution function (including Rician, Nakagami, and Weibull fading, which may be non-identical or correlated, provided that the correlation matrices are of full rank) has been presented in [6]. In particular, it has been shown that full-rank correlation does not affect the DMT, confirming the earlier result in [2] for Rayleigh and Rician channels, and that the DMT is the same in the Rician and Rayleigh channels. Similarly to the original SNR-asymptotic DMT, the results in [6] require $\gamma \to \infty$.

An in-depth study of the DMT in the uncorrelated Rician channel have been presented in [7], confirming the earlier result in [6] that the SNR-asymptotic DMT of Rayleigh and Rician channels are the same. While most of the results in [7] still require $\gamma \to \infty$, a finite-SNR DMT has been presented for SIMO/MISO channels. The SNR-asymptotic DMT of double-scattering MIMO channels under Rayleigh fading has been found in [8], which, in the case of a single double-scattering process (equivalent to the single keyhole channel in [26]), is given by [9]

$$d(r) = \min(m, n)(1 - r), \quad 0 \leq r \leq 1, \quad (8)$$

i.e the DMT curve of the single keyhole channel is significantly lower comparing to that of the full-rank Rayleigh-fading channels (compare (8) to (4): while the maximum diversity gain in (8) is $\min(m, n)$, it is $m \cdot n$ in (4)). It was also elegantly demonstrated in [8], based on asymptotic singular value inequalities, that full-rank correlation does not affect the DMT for *any* fading distribution.

Other directions of active research in this area include the DMT studies of frequency-selective channels [11], of channels with antenna selection [12], of half/full duplex relay channels under various protocols [13]-[15], of ARQ channels [17] and the impact of partial channel state information at the transmitter [16]. Finally, inspired by the DMT framework, a number of space-time coding techniques have been proposed that achieve the diversity-multiplexing tradeoff [18]-[22]. In particular, using an explicit code construction, it has been demonstrated in [19] that the DMT for $T \geq m$ is the same as that with $T \geq m+n-1$, i.e. the minimum number of symbols required to achieve the DMT is $m$ rather than $m + n - 1$.

In the present paper, we adopt a different approach to DMT analysis, which allows us to evaluate a finite-SNR DMT in a closed form for a broad class of fading channels. To evaluate the DMT for arbitrary SNR in a certain channel, one would need to known the outage capacity distribution $F_C(R)$ of that channel. While some results of this kind are available in the literature, their complexity prevents any analytical development, which is the ultimate reason for using $\gamma \to \infty$ in most studies. On the other hand, a number of compact analytical results, which hold at finite SNR, have recently appeared on the outage capacity distribution of asymptotically large systems, i.e. when either $n \to \infty$ or $m \to \infty$, or both [23]-[30]. For a broad class of fading distributions (under mild technical conditions), it turns out to be Gaussian with the mean and the variance determined by the SNR and specifics of the channel.

We exploit these size-asymptotic results to derive the diversity-multiplexing tradeoff at finite SNR and also for a broad class of fading distributions (e.g. not necessarily Rayleigh i.i.d. channels), which we term here "size-asymptotic DMT" to distinguish it from the SNR-asymptotic one in (4). The advantage of this approach is that its results apply at any finite SNR and, thus, have operational significance at realistic SNR values. Furthermore, since the fading distribution is allowed to belong to a broad class rather than being narrowly defined, the results are robust from the practical viewpoint. Since all practical systems operate at finite SNR, the SNR-asymptotic DMT in (4) serves only as an approximation. It is thus important to evaluate the rate of convergence to the limit in (4). Our approach demonstrates that the convergence of the finite-SNR DMT to the SNR-asymptotic one in (4) as the SNR grows is very slow (as $1/(\ln\gamma)^2$) for moderately-large systems and also depends on the system size $n \times m$ and multiplexing gain $r$: while the convergence to the SNR asymptote in (4) is achieved at high but realistic SNR values (e.g. 20dB) for smaller systems (e.g. $2 \times 2$) and large $r$, it is only achieved at unrealistically high SNR (e.g. 80dB) for larger systems (e.g. $10 \times 10$). On the other hand, the size-asymptotic capacity distributions result in compact closed-form approximations of the DMT at realistic SNR values, which are also sufficiently accurate for small systems (e.g. $2 \times 2$)[3].

---

[3] This follows from our results on the size-asymptotic DMT in section III, and also from earlier results in [23]-[30] on the outage capacity distribution under fixed rate (which corresponds to $r = 0$).

The multiplexing gain definition is shown to affect critically the rate of convergence of the finite-SNR DMT to the SNR-asymptotic one: when the multiplexing gain is defined via the mean (ergodic) capacity, the convergence (within reasonable accuracy) takes place at realistic SNR values. Furthermore, in this case the diversity gain can also be used to estimate the outage probability with reasonable accuracy. The multiplexing gain definition via the high-SNR asymptote of the mean capacity (as in [1]) results in very slow rate of convergence for moderate to large systems and, hence, the SNR-asymptotic DMT cannot be used at realistic SNR values. For this definition, the high-SNR threshold required to achieve the DMT in (4) within reasonable accuracy increases exponentially in the number of antennas and in the multiplexing gain. Furthermore, the SNR-asymptotic diversity gain in (3) cannot be used alone to estimate $P_{out}$ in (6) at any SNR (even very large) since the constant $c$, termed "SNR offset", can be very large (e.g. $10^4$) for moderate to large systems. The SNR offset can be somewhat eliminated by proper modifications of (2) and (3), which speed up the convergence in SNR, but the problem still persists. On the contrary, the size-asymptotic approach which we advocate here provides not only the finite-SNR DMT, but also the SNR offset $c$ and thus an accurate estimate of the outage probability (unless the SNR is very high). The finite-SNR diversity gain is shown to decrease with correlation and power imbalance in the channel according to the measure of the latter two introduced in [26], [29], i.e. unlike the SNR-asymptotic DMT, the size-asymptotic DMT adequately describes the outage probability in correlated channels. Furthermore, the effect of correlation and power imbalance on the finite-SNR DMT is described in a compact, closed form for a broad class of fading distributions.

Systems with unequal number of Tx and Rx antennas exhibit qualitatively-different behavior from those with equal number of antennas: while the size-asymptotic DMT of the latter converges to the SNR-asymptotic DMT as the SNR grows, that of the former does not. The size-asymptotic DMT does however provide an accurate approximation of the true DMT at low to moderately-high SNR, even for a modest number of antennas. In this case, the size-asymptotic DMT is complementary to the SNR-asymptotic one: while the latter is accurate at very high SNR, the former is accurate at low to moderately-high SNR. Combining these two, one obtains a DMT estimate that is accurate at the whole SNR range.

Systems/codes are often designed and compared based on their SNR-asymptotic DMT [18]-[22]. However, better DMT does not imply better outage probability at finite SNR, because of the contribution of the SNR offset ignored in the DMT framework. Likewise, equal DMT does not imply equal outage probability. These qualitative observations are substantiated in the paper via a quantitative analysis based on the size-asymptotic theory. The main results are summarized as follows:

• The size-asymptotic, finite-SNR DMT is derived for a broad class of full-rank and rank-deficient fading channels, which is not only accurate at realistic SNR values, but is also an important part in an accurate characterization of the outage probability. (Theorems 4-6, Propositions 3, 4).

• The SNR offset, which is a missing link between the diversity gain and the outage probability, is introduced and characterized via the size-asymptotic theory for a broad class of fading channels. The diversity gain along is shown to be inadequate in characterizing the outage probability. While the multiplexing gain definition via the mean capacity results in a moderate SNR offset, the other definitions (via the high-SNR approximations of the mean capacity) result in a very high SNR offset (Theorem 4, Proposition 3).

• A number of limitations and difficulties of the DMT framework at finite SNR are discussed. This includes a significant impact of the multiplexing gain definition on the finite-SNR DMT (unlike the SNR-asymptotic one), very slow convergence of the finite-SNR DMT to the SNR-asymptotic one, and anomalous behavior of the outage probability at low to moderately high SNR under the DMT framework. (Section III-A).

• Correlation and power imbalance are shown to have a negative impact on the finite-SNR DMT, which is characterized for a broad class of fading channels via the measure of correlation and power imbalance introduced in [26], [29]. (Theorems 5, 6)

• The outage capacity distribution of rank-deficient (double-scattering, multi-keyholes, relay) correlated channels subject to a broad class of fading distributions is obtained in the size-asymptotic regime (Theorem 3).

The rest of the paper is organized as follows. In section II, we introduce the basic system model and briefly review the asymptotic outage capacity distributions (Theorems 1-3). Section III discusses the main limitations and difficulties of the DMT framework and presents the size-asymptotic, finite-SNR DMT and SNR offset for a broad class of full-rank and rank-deficient fading channels (Theorems 4-6, Propositions 3, 4). Finally, section IV concludes the paper. The proofs are collected in Appendix.

## II. SYSTEM MODEL AND OUTAGE CAPACITY DISTRIBUTION

The standard baseband discrete-time system model is adopted here,

$$\mathbf{r} = \mathbf{H}\mathbf{s} + \boldsymbol{\xi}, \qquad (9)$$

where $\mathbf{s}$ and $\mathbf{r}$ are the Tx and Rx vectors correspondingly, $\mathbf{H}$ is the $n \times m$ frequency-flat, block-fading channel matrix, i.e. the matrix of the complex channel gains between each Tx and each Rx antenna, $m$ and $n$ are the numbers of Tx and Rx antennas, and $\boldsymbol{\xi}$ is the additive white Gaussian noise (AWGN), which is assumed to be $\mathcal{CN}(\mathbf{0}, \sigma_0^2 \mathbf{I})$, i.e. independent and identically distributed (i.i.d.) in each branch. The assumptions on the distribution of $\mathbf{H}$ follow those of the asymptotic capacity distributions (discussed below): the entries of $\mathbf{H}$ are assumed to be either (i) i.i.d. but otherwise arbitrary fading (this includes Rayleigh fading as a special case) [27], (ii) unitary-independent-unitary (UIU) [28], (iii) correlated Rayleigh-fading with separable correlation structure [24], [30], or (iv) follow the statistics of the correlated double-scattering or keyhole channel [26], [29].

When full channel state information (CSI) is available at the Rx end but no CSI at the Tx end, the instantaneous channel capacity (i.e. the capacity of a given channel realization $\mathbf{H}$) in nats/s/Hz is given by the celebrated log-det formula [31], [32],

$$C = \ln \det \left( \mathbf{I} + \frac{\gamma}{m} \mathbf{H}\mathbf{H}^+ \right), \quad (10)$$

where $\gamma$ is the average SNR per Rx antenna (contributed by all Tx antennas), "+" denotes conjugate transpose.

For large $m, n$, the distribution of $C$ takes on a remarkably simple form in a number of cases[4]:

*Theorem 1:* Let $\mathbf{H}$ be an $n \times m$ channel matrix whose entries are i.i.d. zero mean random variables with unit variance and $E|H_{ij}|^4 = 2$. As $m, n \to \infty$ and $\beta = m/n$ is a constant, the instantaneous capacity in (10) is asymptotically (in $m, n$) Gaussian, with the following mean $\overline{C}$ and variance $\sigma_C^2$:

$$\frac{\overline{C}}{n} = \beta \ln \left( 1 + \frac{\gamma}{\beta} - \frac{1}{4} F\left(\frac{\gamma}{\beta}, \beta\right) \right)$$
$$+ \ln \left( 1 + \gamma - \frac{1}{4} F\left(\frac{\gamma}{\beta}, \beta\right) \right) - \frac{\beta}{4\gamma} F\left(\frac{\gamma}{\beta}, \beta\right), \quad (11)$$

$$\sigma_C^2 = -\ln \left( 1 - \beta \left[ \frac{1}{4\gamma} F\left(\frac{\gamma}{\beta}, \beta\right) \right]^2 \right), \quad (12)$$

where $F(x, z) = (\sqrt{x(1+\sqrt{z})^2 + 1} - \sqrt{x(1-\sqrt{z})^2 + 1})^2$.
*Proof:* see [[27], Theorem 2.76]. ■

Moreover, from [[28], Theorem 5], the instantaneous capacity of a channel whose channel matrix has independent but not necessarily identically distributed entries is also asymptotically Gaussian, as both $m, n \to \infty$, with the mean and variance defined by (11) and (12) respectively, if the channel gain matrix $E|h_{ij}|^2$ is asymptotically mean double-regular (see [[28], Definition 3]). This implies that the instantaneous capacity of a broad class of so-called unitary-independent-unitary (UIU) channels [28] is also asymptotically Gaussian.

At moderate to high SNR, (11), (12) can be approximated as[5]

$$\overline{C} \approx \min(m, n) \ln\left(\frac{\gamma}{a}\right),$$
$$\sigma_C^2 \approx \begin{cases} -\ln(1-\beta), & \beta < 1 \\ \frac{1}{2}\left(\ln\frac{\gamma}{4} + \frac{2}{\sqrt{\gamma}}\right), & \beta = 1 \\ -\ln(1-1/\beta), & \beta > 1 \end{cases}, \quad (13)$$

where $a$ is the SNR offset,

$$a = \begin{cases} e\beta(1-\beta)^{1/\beta - 1}, & \beta < 1 \\ e, & \beta = 1 \\ e(1-1/\beta)^{\beta-1}, & \beta > 1 \end{cases} \quad (14)$$

Note that Theorem 1 applies to a broad class of channels, not only Rayleigh or Rician ones (only these channels were considered in [1], [2]), and also includes the channels not considered in [6].

[4] Other asymptotic results are also available in the literature. However, we will rely only on these theorems in the present paper.
[5] Similar approximations, but without $1/\sqrt{\gamma}$ term, can be found elsewhere in the literature. They, however, become accurate for significantly larger SNR, $\gamma \geq 20...30 dB$, while the approximation in (13) is already accurate at $\gamma \geq 5 dB$.

*Theorem 2:* Let $\mathbf{H}$ be an $n \times m$ matrix of a Rayleigh-fading channel with separable (Kronecker) correlation structure, such that $E(vec(\mathbf{H})vec(\mathbf{H}^+)) = \mathbf{R}_t^T \otimes \mathbf{R}_r$, where $\mathbf{R}_t = n^{-1} E(\mathbf{H}^+\mathbf{H})$, $\mathbf{R}_r = m^{-1} E(\mathbf{H}\mathbf{H}^+)$ are transmit and receive correlation matrices respectively, operator $vec(\mathbf{H})$ creates a column vector by stacking the elements of $\mathbf{H}$ columnwise, $\otimes$ is the Kronecker product, and $\mathbf{R}_t^T$ is the transpose of $\mathbf{R}_t$. Then, as $m \to \infty$, and

$$\lim_{m \to \infty} \|\mathbf{R}_t\|_2 / \|\mathbf{R}_t\| = 0, \quad (15)$$

where $\|\ \|_2$ and $\|\ \|$ are spectral and Frobenius norms respectively, the instantaneous capacity in (10) is asymptotically Gaussian with the following mean $\overline{C}$ and variance $\sigma_C^2$

$$\overline{C} = \ln \det (\mathbf{I} + \gamma \mathbf{R}_r) \approx n \ln\left(\frac{\gamma}{a}\right), \quad (16)$$

$$\sigma_C^2 = \frac{1}{m^2} \|\mathbf{R}_t\|^2 \cdot \sum_{k=1}^n \left(\frac{\gamma \lambda_k^r}{1 + \gamma \lambda_k^r}\right)^2 \approx \frac{n}{m^2} \|\mathbf{R}_t\|^2, \quad (17)$$

where $\lambda_k^r$, $k = 1...n$, are the eigenvalues of $\mathbf{R}_r$, $a = (\det \mathbf{R}_r)^{-1/n}$ is the SNR offset, and the approximation holds at high SNR, when $\gamma \lambda_{\min}^r \gg 1$, under the normalization $tr\mathbf{R}_r = n$.
*Proof:* see [24], [25]. ■

When both $m, n \to \infty$ and $\beta = m/n$ is a constant, it has been shown that under certain general conditions on the channel correlation, the instantaneous capacity of a Rayleigh-fading channel with the Kronecker correlation structure is also asymptotically Gaussian [30]. Note that the UIU channels considered in [[28], Theorem 5] and those in [30] do not overlap, unless $\mathbf{R}_t = \mathbf{I}$, $\mathbf{R}_r = \mathbf{I}$ (uncorrelated case).

Theorems 1 and 2 apply to full-rank channels. Rank-deficient channels can be considered via the multi-keyhole model in [26], [29]. Using [[26], Theorems 4, 7], Comment 5 in [25], and Von-Neumann trace inequality [37], the following theorem follows.

*Theorem 3:* Consider a rank-deficient channel of the form $\mathbf{H} = \sum_{k=1}^M b_k \mathbf{h}_{rk} \mathbf{h}_{tk}^+$, $rank(\mathbf{H}) = M \leq \min(m, n)$, where $b_k$ are complex modal amplitudes, $\mathbf{h}_{tk}$ and $\mathbf{h}_{rk}$ are modal Tx and Rx channel vectors, which are independent of each other, with correlation matrices $\mathbf{R}_{tk} = E\{\mathbf{h}_{tk}\mathbf{h}_{tk}^+\}$, $\mathbf{R}_{rk} = E\{\mathbf{h}_{rk}\mathbf{h}_{rk}^+\}$. Assume the following conditions hold:

(a) $\mathbf{h}_{tk} = \mathbf{R}_{tk}^{1/2} \mathbf{g}_{tk}$, $\mathbf{h}_{rk} = \mathbf{R}_{rk}^{1/2} \mathbf{g}_{rk}$, where $\mathbf{g}_{tk}$ and $\mathbf{g}_{rk}$ are zero mean, unit variance, complex circular symmetric random vectors with i.i.d. entries (not necessarily Gaussian),

(b) $m_{2+\delta}(|g|^2) < \infty$ for some $\delta > 0$ and $m_2(|g|^2) > 0$, where $m_\delta(x) = E(x - Ex)^\delta$ is the central moment of $x$ of order $\delta$, and $g$ is any entry of $\mathbf{g}_{tk}$ or $\mathbf{g}_{rk}$,

$$(c) \quad \lim_{m \to \infty} \frac{\|\mathbf{R}_{tk}\|_2}{\|\mathbf{R}_{tk}\|} = \lim_{n \to \infty} \frac{\|\mathbf{R}_{rk}\|_2}{\|\mathbf{R}_{rk}\|} = 0, \quad \forall k.$$

As $m, n \to \infty$, the instantaneous capacity of this channel is asymptotically Gaussian with the following mean and vari-ance,

$$\overline{C} = \sum_{k=1}^{M} \ln\left(1 + |b_k|^2 \gamma\right) \approx M \ln\left(\frac{\gamma}{a}\right);$$

$$\sigma_C^2 = \sum_{k=1}^{M} \left(\frac{|b_k|^2 \gamma}{1 + |b_k|^2 \gamma}\right)^2 \left(\frac{\beta_{tk}}{m^2} \|\mathbf{R}_{tk}\|^2 + \frac{\beta_{rk}}{n^2} \|\mathbf{R}_{rk}\|^2\right)$$

$$\approx \sum_{k=1}^{M} \left(\frac{\beta_{tk}}{m^2} \|\mathbf{R}_{tk}\|^2 + \frac{\beta_{rk}}{n^2} \|\mathbf{R}_{rk}\|^2\right), \quad (18)$$

where $\gamma$ is the total SNR (combined from all Rx antennas), $a = \prod_k |b_k|^{-2/M}$ is the SNR offset, $\beta_{tk} = m_2(|g_{tk}|^2)$, $\beta_{rk} = m_2(|g_{rk}|^2)$, and $g_{t(r)k}$ is any entry of $\mathbf{g}_{t(r)k}$. The approximation holds at moderate to high SNR, when $\gamma |b_{\min}|^2 \gg 1$. If $\mathbf{g}_{tk}$, $\mathbf{g}_{tk}$ are Gaussian, then $\beta_{tk} = \beta_{rk} = 1$.

*Proof:* see Appendix. ∎

It should be noted that the channel model in Theorem 3 coincides with the amplify-and-forward relay channel when each relay node has a single antenna and the relay noise is negligible (see [41] for examples of such cases). Therefore, under this condition, the relay channel will have the same outage capacity and the DMT as the multi-keyhole one. Additionally, from [[29], Theorem 2], the channel in Theorem 3 converges to a Rayleigh-fading one as $M \to \infty$, whose outage capacity is also asymptotically Gaussian [30]. The condition (c) in Theorem 3 implies that the channel is "asymptotically uncorrelated" in the sense that the measure of correlation and power imbalance approaches zero [[25], Comment 5].

Using the asymptotic outage capacity distributions above, the outage probability can now be compactly expressed as

$$P_{out}(R) = Q\left(\frac{\overline{C} - R}{\sigma_C}\right) \quad (19)$$

where $Q(x) = \frac{1}{\sqrt{2\pi}} \int_x^\infty \exp(-t^2/2)dt$. It follows from (19) that reliable transmission (low $P_{out}$) is possible when $R < \overline{C} - \sigma_C$ and, for a given rate, the larger the mean capacity $\overline{C}$ and the smaller the variance $\sigma_C^2$, the smaller $P_{out}$ is. In the following sections, we exploit (19) to evaluate the finite-SNR DMT.

## III. FINITE-SNR DMT VIA SIZE-ASYMPTOTIC CAPACITY DISTRIBUTION

We begin with a motivation of the size-asymptotic analysis as an alternative to the SNR-asymptotic one by pointing out some limitations and difficulties of the DMT framework when applied to finite (realistic) SNR values.

### A. Limitations and difficulties of the DMT framework at finite SNR

*Multiplexing gain definitions:* While a finite-SNR DMT analysis requires using finite-SNR analogs of the definitions in (2), (3), their straightforward extensions, i.e.

$$r = \frac{R}{\ln \gamma}, \quad d_\gamma = -\frac{\ln P_{out}}{\ln \gamma}, \quad (20)$$

produce a number of difficulties pointed out below. In particular, the convergence of the finite-SNR DMT to the asymptotic one in (4) as the SNR grows is very slow and can be significantly improved if $r$ is defined via $\overline{C}$, or via $\ln(\gamma/a)$, which is motivated by (13) and takes into account the SNR offset[6] $a$,

$$r = \frac{\min(m,n)R}{\overline{C}} \quad (21)$$

$$r = \frac{R}{\ln(\gamma/a)} \quad (22)$$

where (21) defines the rate via the mean capacity per degree of freedom, $R = r\overline{C}/\min(m,n)$. Note that, at finite SNR, $r = 0$ corresponds to $R = 0$. While the SNR-asymptotic DMT is the same for all 3 definitions of the multiplexing gain, there is a significant difference at finite SNR, both in terms of diversity gain and SNR offset. The difference in diversity gains does not disappear unless the SNR is unreasonably high, and the difference in SNR offsets does not disappear at any SNR, does not matter how high. This motivates the study of all 3 definitions to select the best one at finite SNR.

*Diversity gain and SNR offset:* Another finite-SNR difficulty is that when $P_{out}$ behaves as in (6), which serves as a baseline model for the finite-SNR DMT analysis, the finite-SNR diversity gain $d_\gamma = -\ln c/\ln \gamma + d$ includes the effect of the SNR offset $c$ and is not equal to the "true" diversity gain $d$, unless $\ln \gamma$ is very (unrealistically) high. The difference between $d$ and $d_\gamma$ can be significant when the SNR offset $c$ is significant (i.e. either too high or too low). A definition of the finite-SNR diversity gain introduced in [2] partially eliminates this problem and captures the differential effect of diversity, i.e. how much increase in SNR is required to decrease $P_{out}$ by certain amount,

$$d'_\gamma = -\frac{\partial \ln P_{out}}{\partial \ln \gamma} \quad (23)$$

When $P_{out}$ is as in (6) and $c, d$ are SNR independent, $d'_\gamma = d$, i.e. this definition recovers precisely the "true" diversity gain at finite SNR. Furthermore, since the differential diversity gain $d'_\gamma$ is insensitive to the SNR offset $c$ in (6), the convergence to the SNR-asymptotic value is faster. For sufficiently high SNR, both definitions of the diversity gain (in (23) and (20)) give similar results and, when the limit exists, $\lim_{\gamma \to \infty} d_\gamma = \lim_{\gamma \to \infty} d'_\gamma$ [7]. However, $P_{out}$ cannot be reliably estimated from $d'_\gamma$ alone since it captures only the differential effect of increasing SNR and is independent of the offset $c$, which may significantly affect $P_{out}$ (see, for example, Fig. 1). Motivated by (6), we define the SNR offset for given $P_{out}$ as

$$c_\gamma = P_{out} \gamma^{d'_\gamma} \quad (24)$$

On the other hand, given both the diversity gain and the SNR offset, the outage probability can be estimated from (6). Thus, the SNR offset provides the missing link between the DMT and outage probability[7], and also indicates how far away the rough estimate $P_{out} \approx 1/\gamma^d$ is. Using the size-asymptotic results above, the SNR offset can be evaluated with sufficient accuracy, even for small to moderate-size systems, as demonstrated below.

---
[6] [33] gives a detailed discussion of the importance of SNR offset in the capacity analysis of MIMO systems. Note that this offset is missing in (5).

[7] while, for most channels at finite SNR, $d'_\gamma$ and $c_\gamma$ are not SNR-independent constants but rather slowly-varying functions of the SNR, the DMT framework can still be used.

*Outage probability in the DMT framework:* While the diversity gain provides some indication of the performance, its usefulness lies in its relation with the outage probability (or the average error rate) as the latter is the ultimate performance indicator, not the diversity gain itself. To demonstrate the impact of multiplexing gain definitions and to test the suitability of size-asymptotic capacity distribution to predict the outage probability of finite-size systems under the DMT framework, Fig. 1 and 2 compare the outage probability vs. SNR from the asymptotic result in (19) (evaluated based on $\overline{C}$ and $\sigma_C^2$ given by Theorem 1) to Monte-Carlo (MC) simulations for i.i.d. Rayleigh-fading channel using the multiplexing gain definitions in (20)-(22). A good agreement between the size-asymptotic and MC results is observed even for small system size $m = n = 2$, demonstrating that the size-asymptotic theory is practically relevant. Fig. 1, 2 also demonstrate a limitation of the DMT framework with the multiplexing gain definitions in (20) and (22), which is the anomalous behavior of the outage probability (increasing with the SNR) for low to moderate SNR range. This is due to the fact that the rate $R < \overline{C}$ on the corresponding interval, but it increases faster than $\overline{C}$ with the SNR, so that $|\overline{C} - R|/\sigma_C$ decreases; after the anomalous region this tendency is reversed. This never happens if multiplexing gain is defined via the mean capacity as in (21). Also note a high SNR offset ($c \approx 10^4$, see (6)) in $P_{out}$ for $R = r \ln \gamma$ and $n = 10$. This makes it impossible to estimate $P_{out}$ from the diversity gain alone, i.e. using $P_{out} \approx 1/\gamma^d$ (as suggested in [35]), no matter how high the SNR is. The rough estimation $P_{out} \approx 1/\gamma^d$ works only if $c$ is on the order of unity. When this is not the case $c$ has to be accounted for as well.

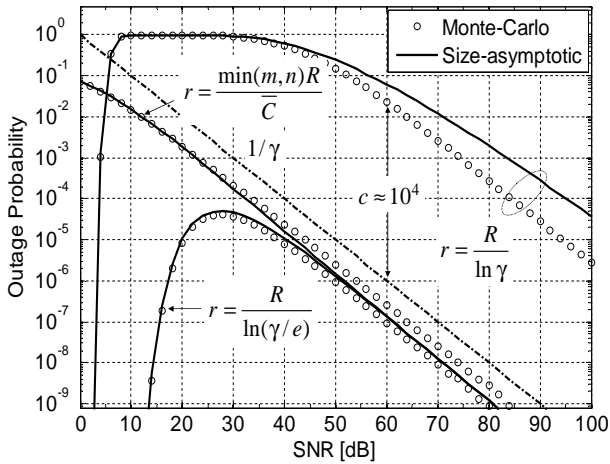

Fig. 1. Outage probability vs. SNR for various definitions of the multiplexing gain; $n = m = 10, r = 9, d(r) = 1$; solid line - asymptotic from (11), (12), (19), circles - Monte-Carlo simulations ($10^{10}$ trials); dashed line - $P_{out} = 1/\gamma$. Note high SNR offset ($c \approx 10^4$). Asymptotic approximation is accurate for all multiplexing gain definitions but (20) at very high SNR.

The following proposition formalizes this limitation of the DMT framework for a broad class of fading channels.

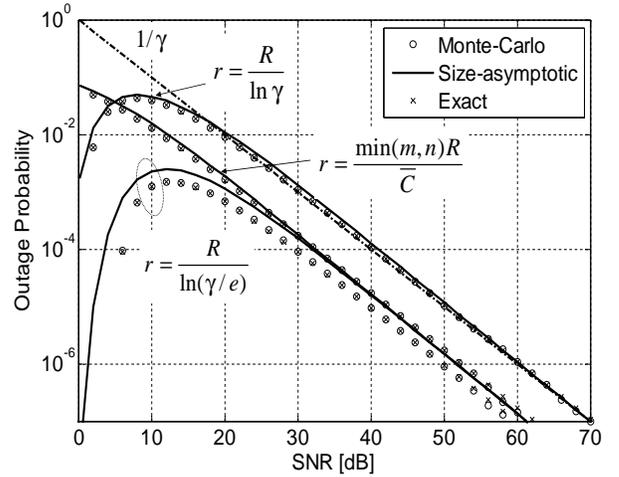

Fig. 2. Outage probability vs. SNR for various definitions of the multiplexing gain; $n = m = 2, r = 1, d(r) = 1$; solid line - asymptotic from (11), (12), (19), circles - Monte-Carlo simulations ($10^8$ trials); dashed line - $P_{out} = 1/\gamma$. The SNR offset is small in this case ($c \approx 1$) and the convergence is achieved at realistic SNR. Asymptotic approximation is accurate for all multiplexing gain definitions. The exact outage is obtained by integration of the Wishart eigenvalue density (see e.g. [31]).

*Proposition 1:* Consider a finite SNR DMT, such that $d(r) < \infty$ as $r \to 0$. Assume that $\lim_{\varepsilon \to 0} F_C(\varepsilon) = 0$ and that the outage probability is as in (6). Then, $c(r) \to 0$ as $r \to 0$, i.e. unbounded SNR offset for small $r$.

*Proof:* see Appendix. ■

Majority of known channels satisfy the conditions of Proposition 1, i.e. the diversity gain is bounded for small $r$, and the probability of zero channel capacity is zero. Proposition 1 implies that when $r$ is small, there always is a very significant SNR offset under the model in (6). We conclude that when comparing two systems, $d_1 > d_2$ does not imply $P_{out,1} < P_{out,2}$ at finite SNR, since it may be that $c_1 > c_2$ and the latter effect is dominant. Likewise, $d_1 = d_2$ does not imply $P_{out,1} = P_{out,2}$ at any SNR, unless $c_1 = c_2$. Hence, using the DMT curves alone to compare two systems may produce incorrect results, even at very high SNR. This suggests that the SNR offset $c$ should also be included in the DMT framework. Fig. 3 shows the offset $c(r)$ vs. $r$ evaluated for a $2 \times 2$ i.i.d. Rayleigh-fading channel at $\gamma = 10dB$. It follows that $c \approx 1$ at $r > 1.5$ and the rough estimate $P_{out} \approx 1/\gamma^{d(r)}$ is reasonably accurate. However, $c$ rapidly decreases at $r < 1$ and it is impossible to estimate $P_{out}$ from the rough approximation above in this range. This observation may have significant consequences for the design of DMT-achieving codes (see [18]-[22] for examples of such designs).

The problem of significant SNR offset is somewhat eliminated, for moderate to high $r$, by using the multiplexing gain definition in (22), as $c$ becomes a moderate constant, but the anomalous behavior of the outage probability is not eliminated so that its estimation from the diversity gain alone at $\gamma \leq 30dB$ is not possible. Using the definition in (21) eliminates most of the problem, leaving only the moderate offset $c \approx 1/5$. For smaller systems (Fig. 2), this problem is not that severe and the SNR offset disappears at $\gamma \geq 15dB$, but the anomalous behavior of the outage probability at low to moderate SNR for all definitions of the multiplexing gain but in (21) is still present.

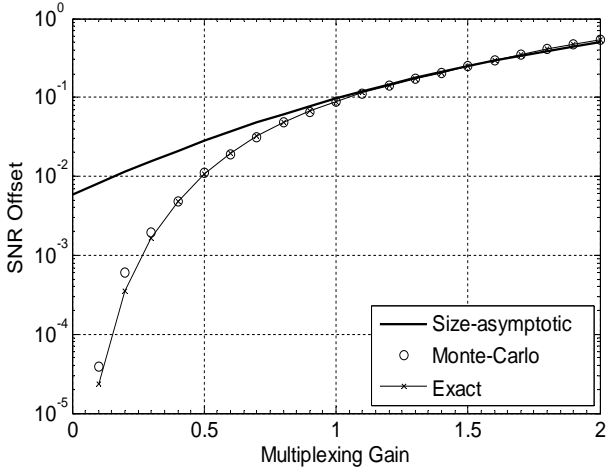

Fig. 3. SNR offset vs. multiplexing gain (21) in $2 \times 2$ i.i.d. Rayleigh fading channel at SNR=10dB. While the offset is not significant at $r > 1.5$, it rapidly becomes significant at $r < 1$ so that the rough estimate $P_{out} \approx 1/\gamma^{d(r)}$ is not accurate anymore. The Gaussian approximation is not accurate when $r \to 0$, but is accurate when multiplexing gain is moderate to high, which is a practically-important range for modern systems (see e.g. [40]).

### B. Size-asymptotic DMT and SNR offset

In this section, the finite-SNR DMT is analyzed via the size-asymptotic capacity distribution in (19) under the multiplexing gain definitions in (20)-(22) to show their advantages and disadvantages when applied to realistic systems (low to moderate SNR, moderate or small system size). The following proposition is instrumental in using the generic Gaussian capacity distribution for the size-asymptotic DMT analysis.

*Proposition 2:* The size-asymptotic DMT under the outage capacity distribution in (19) is

$$d_\gamma \approx \frac{1}{2\ln\gamma}\left(\frac{\overline{C}_1}{\sigma_C}\right)^2 d^*(r), \qquad (25)$$

$$d'_\gamma \approx \frac{1}{2}\frac{\partial}{\partial\ln\gamma}\left(\frac{\overline{C}_1}{\sigma_C}\right)^2 d^*(r) \qquad (26)$$

where $\overline{C}_1 = \overline{C}/m^*$ is the mean capacity per degree of freedom in the channel, $m^* = min(m,n,M)$ is the channel rank, $d^*(r) = (m^* - r)^2$, and $r$ is the multiplexing gain defined via the mean capacity in (21).

*Proof:* see Appendix. ∎

Note that $d^*(r) = (m^* - r)^2$ is somewhat similar to the original Zheng-Tse DMT in (4), but also has two notable differences: (i) for a full-rank channel, it depends only on $\min(m,n,M)$, not on $m,n,M$ individually, and coincides with (4) when $m = n = M$; (ii) there is no linear interpolation for non-integer $r$.[8] While $d^*(r)$ is independent of the SNR, the first two factors in (25) describe the effect of SNR and

[8]this also holds true for most fading channels at finite SNR [38].

also of channel correlation. As will be demonstrated below, under certain circumstances $\lim_{\gamma \to \infty} d_\gamma = d^*(r)$, so that the size-asymptotic DMT converges to the SNR-asymptotic one for integer $r$.

To simplify the analysis and to get some insight, we use below high but finite SNR approximations, i.e. $\gamma >> 1$, but not $\gamma \to \infty$. These approximations, as it is demonstrated below, hold true already at low or moderate SNR levels and allow one to quantify the effect of SNR on the DMT and, in particular, to establish the SNR levels at which the asymptotic results in [1] are sufficiently accurate.

*Theorem 4:* Consider a full-rank, i.i.d., $n \times n$ arbitrary fading channel under the conditions of Theorem 1. Its size-asymptotic finite-SNR DMT can be approximated as

$$d'_\gamma \approx (n-r)^2\left(1 - \frac{1}{2\sqrt{\gamma}}\right), \quad \gamma \geq 1, \qquad (27)$$

where the multiplexing gain $r = nR/\overline{C}$ is defined via the mean capacity. The SNR offset $c_\gamma$ is

$$c_\gamma \approx \frac{e^{d_\gamma}}{\sqrt{4\pi d_\gamma \ln(\gamma/e)}} \qquad (28)$$

$$\to \frac{e^{d(r)}}{\sqrt{4\pi d(r)\ln(\gamma/e)}}, \text{ as } \gamma \to \infty, \qquad (29)$$

where $d_\gamma = (n-r)^2(1 + 2/[\sqrt{\gamma}\ln(\gamma/e)])$ and the approximation in (28) holds for $d_\gamma \ln(\gamma/e) > 1$ and $0 < r < n$; $c_\gamma \to 0$ when $r \to 0$, and $d_\gamma = 0$, $c_\gamma = 1/2$ for $r = n$. The outage probability can be estimated as $P_{out} \approx c_\gamma/\gamma^{d'_\gamma}$.

*Proof:* see Appendix. ∎

Note that the first factor in (27) is identical to the SNR-asymptotic DMT in (4) (except for missing linear interpolation), and the second term represents the effect of the finite SNR. The i.i.d. Rayleigh channel considered in [1] and Rician one considered in [2] are special cases of Theorem 4. Note also that $\lim_{\gamma \to \infty} d_\gamma = \lim_{\gamma \to \infty} d'_\gamma = d(r)$ and the convergence of $d'_\gamma$ to the SNR-asymptotic $d(r)$ in (4) takes place when the second term in (27) can be neglected, which we set, somewhat arbitrary, as $1/(2\sqrt{\gamma}) \leq 0.1$ (i.e. within 10% accuracy), so that

$$d'_\gamma \approx d(r) = (n-r)^2, \text{ for } \gamma \geq 25 \approx 14dB \qquad (30)$$

The following proposition shows that, unlike the SNR-asymptotic DMT, the finite-SNR one depends crucially on the definition of multiplexing gain, which also has a significant effect on the SNR offset.

*Proposition 3:* The size-asymptotic DMT of a full-rank i.i.d. $n \times n$ channel under conditions of Theorem 1 with the approximations in (13) and under the multiplexing gain definitions in (20), (22), is

$$d'_\gamma \approx (n-r)^2\left(1 - \frac{n+r}{n-r}\frac{1}{\sqrt{\gamma}} - \left(\frac{r}{n-r}\right)^2 \frac{1}{\ln(\gamma/e)^2}\right),$$
$$\text{for } r = \frac{R}{\ln\gamma} < n \qquad (31)$$

$$d'_\gamma \approx (n-r)^2\left(1 - \frac{n+r}{n-r}\frac{1}{\sqrt{\gamma}}\right), \text{ for } r = \frac{R}{\ln(\gamma/e)} < n \qquad (32)$$

If $r = n$, then $d_\gamma = d'_\gamma = 0$. The SNR offset is given by

$$c_\gamma \to \frac{e^{d(r)} e^{2r(n-r)}}{\sqrt{4\pi d(r) \ln(\gamma/e)}}, \text{ as } \gamma \to \infty, \text{ for } r = R/\ln\gamma \quad (33)$$

$$c_\gamma \to \frac{e^{d(r)}}{\sqrt{4\pi d(r) \ln(\gamma/e)}}, \text{ as } \gamma \to \infty, \text{ for } r = R/\ln(\gamma/e) \quad (34)$$

*Proof:* along the same line as for Theorem 4. ∎

It follows from Theorem 4 and Proposition 3 that $d'_\gamma \to d_\gamma \to d(r)$ (without linear interpolation) as $\gamma \to \infty$ for all 3 multiplexing gain definitions, but the convergence rate is the fastest for $r = nR/\overline{C}$ and the slowest for $r = R/\ln\gamma$: $d'_\gamma \approx d(r) = (n-r)^2$ at high SNR such that

$$\gamma \geq \max\left[\left(\frac{10(n+r)}{n-r}\right)^2, \exp\left(1 + \frac{3r}{n-r}\right)\right],$$
$$\text{for } r = \frac{R}{\ln\gamma}, \quad (35)$$

$$\gamma \geq \left(\frac{10(n+r)}{n-r}\right)^2, \text{ for } r = \frac{R}{\ln(\gamma/e)}. \quad (36)$$

For moderate to large system size only $r = nR/\overline{C}$ results in the convergence at realistic SNR values (see Fig. 4), so that the SNR-asymptotic DMT has operational significance only for this multiplexing gain definition. Comparing (33) and (34) to (29), one concludes that the SNR offsets for $r = R/\ln(\gamma/e)$ and $r = nR/\overline{C}$ are the same, but there is an additional SNR offset factor $e^{2r(n-r)}$ for $r = R/\ln\gamma$, which can be very significant for large $n$ or $r$, as examples below demonstrate. Based on (29), (33) and (34), we remark that the SNR offset $c_\gamma$ is exponentially large in the diversity gain $d(r)$ for various multiplexing gain definitions. While the $\ln(\gamma/e)$ term somewhat reduces the offset, it is a minor effect since $\sqrt{\ln(\gamma/e)}$ increases very slowly with the SNR.

Figs. 4 and 5 compare the differential diversity gain evaluated via the asymptotic distribution with the moments in (11), (12) to the approximations in (27), (31) and (32), and Fig. 6 does the same for the SNR offset. Few observations are in order:

• The size-asymptotic analysis provides reasonable accuracy in estimating both the diversity gain and the SNR offset, even for small systems.

• The original multiplexing gain definition in (20), which was used in [1], results in extremely slow convergence (as $(\ln\gamma)^{-2}$) of the finite-SNR DMT to the SNR-asymptotic one for large systems (see Fig. 4 and also Fig. 7), making the results inapplicable at realistic SNR values. The SNR offset is very high in this case, $c \approx 10^4$ at $\gamma = 60dB$ for $n = m = 10$, $r = 9$ (see Fig. 1 and (33)), which makes the rough approximation $P_{out} \approx 1/\gamma^d$ inaccurate at any SNR. The high-SNR threshold increases exponentially in system size and in the multiplexing gain (see (35)).

• The high-SNR offset in (22) improves the convergence, but yet not enough to achieve it at realistic SNR for large systems.

• The multiplexing gain definition via the mean capacity in (21) is the best, with the convergence at realistic SNR values, which is also independent of any system parameters, unlike those in (20) and (22).

• Comparing Figs. 4 and 5, one concludes that the convergence of the finite-SNR DMT to the SNR-asymptotic one for the multiplexing gains in (20) and (22) is significantly affected by the system size: for small systems, all three definitions give roughly the same (fast) convergence, achieved at realistic SNRs; for large systems, only the definition in (21) results in convergence at realistic SNRs.

• Unlike large systems (see Fig. 1), the SNR offset for smaller systems (see Fig. 2) is moderate for all multiplexing gain definitions, so that the rough approximation $P_{out} \approx 1/\gamma^d$ can be used, unless $r \to 0$, as indicated in Proposition 1. In the latter case, the rough approximation cannot be used regardless of the multiplexing gain definition and the system size, since, for the baseline model in (6), $c \to 0$ and this is the dominant effect, which makes the DMT framework inapplicable in this case.

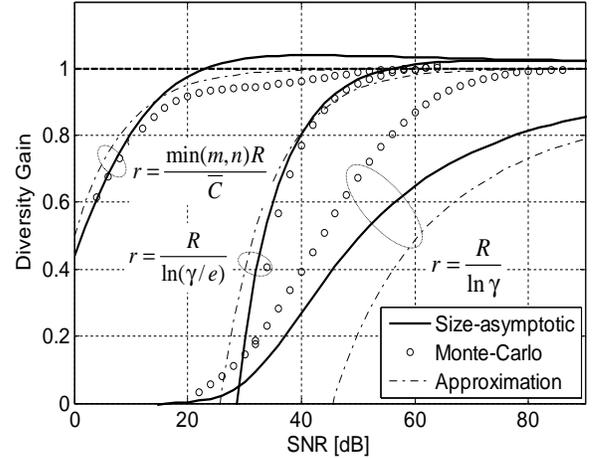

Fig. 4. Differential diversity gain vs. SNR for various definitions of the multiplexing gain; $n = m = 10, r = 9, d(r) = 1$; solid line – asymptotic from (11), (12), (19), dashed – approximations in (27), (31), (32). Convergence to the asymptotic result in (4) is achieved at $\gamma \geq 45dB$ and $\gamma \geq 65dB$ for the multiplexing gain definitions in (22) and (20), respectively, and at $\gamma \geq 14dB$ for that in (21), so only the latter has operational significance at realistic SNR.

While the results above have been obtained for $n = m$ channels, similar results also hold for $m \neq n$ channels, which is briefly summarized below. A more detailed discussion can be found in [38].

*Proposition 4:* Under the conditions of Theorem 1, the size-asymptotic DMT of a full rank $n \times m$ channel, $n \neq m$, is given by

$$d'_\gamma \approx 2d_\gamma \approx (m^* - r)^2 \frac{\ln(\gamma/a)}{-\ln(1-\beta^*)}, \quad 0 \leq r \leq m^*, \quad (37)$$

where $r = m^* R/\overline{C}$, $m^* = \min(m,n)$, and $\beta^* = \min(m,n)/\max(m,n)$.

*Proof:* see Appendix A. ∎

Note that the first term in (37) is somewhat similar to the SNR-asymptotic DMT of Zheng and Tse in (4), but is affected only by $\min(m,n)$, not $m$ and $n$ individually, and no linear interpolation is present. The size-asymptotic DMT here does not converge to the SNR-asymptotic one as $\gamma \to \infty$, which is due to the fact that the accuracy of the Gaussian approximation for finite-size systems decreases as one moves to the distribution tail [38]. Yet, the approximation in (37) is more accurate than the SNR asymptotic one in (4) for low to moderate SNR range, as Fig. 7 demonstrates. Based on this, we observe that the size-asymptotic and SNR-asymptotic results are complementary in this case: while the latter is more accurate at very high SNR, the former is better at low to moderately high SNR, so that the DMT can be approximated for the whole SNR range as

$$d = \min\{d'_\gamma, d(r)\} \qquad (38)$$

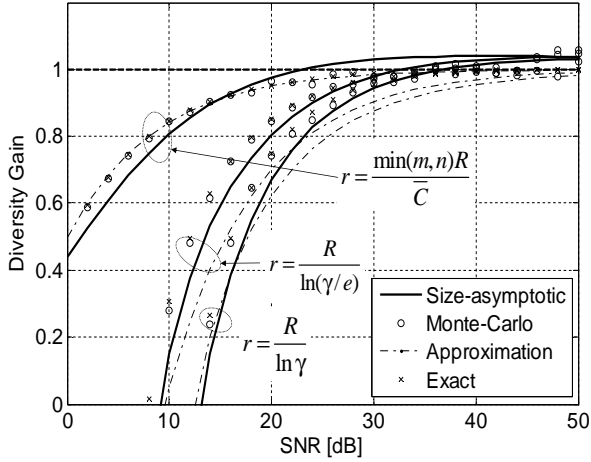

Fig. 5. Differential diversity gain vs. SNR for various definitions of the multiplexing gain; $n = m = 2, r = 1, d(r) = 1$. Convergence to the asymptotic result in (4) is achieved at $\gamma \geq 25dB$ for the multiplexing gain definitions in (20) and (22), and at $\gamma \geq 14dB$ for that in (21), i.e. faster convergence for smaller systems.

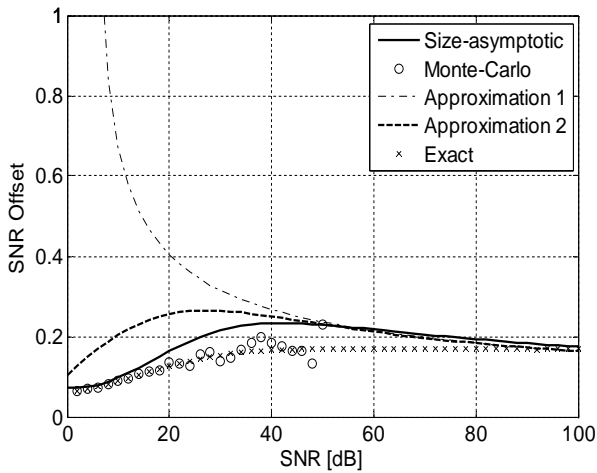

Fig. 6. SNR offset $c_\gamma$ vs. SNR for $2 \times 2$ system and $r = nR/\overline{C} = 1$; Approximations 1 and 2 are as in (29) and (28). The SNR offset is a slowly-varying function of the SNR, which converges to the asymptotic one in (29) at high SNR. The size-asymptotic model provides a reasonable approximation, even for small systems, over the whole SNR range. The exact offset was obtained via the exact outage probability obtained by integration of the Wishart eigenvalue density.

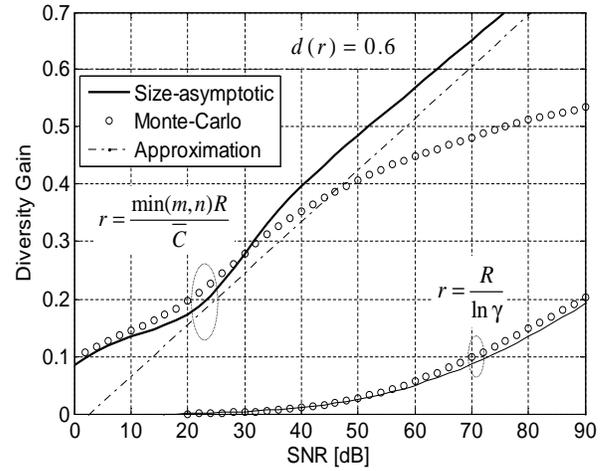

Fig. 7. Differential diversity gain vs. SNR for various definitions of the multiplexing gain; $n = 10, m = 9, r = 8.7, d(r) = 0.6$; solid line – size-asymptotic from (11), (12), (19), dashed – approximation in (37). The size-asymptotic diversity gain is accurate up to about 40dB, and the multiplexing gain definition via the mean capacity is the best one. Similar results can also be observed for smaller-size channels.

### C. The impact of correlation and power imbalance

While all the results above apply to independent channel, similar results can also be obtained for correlated ones based on Theorem 2.

*Theorem 5:* Under the conditions of Theorem 2, the size-asymptotic DMT of a correlated, full-rank $n \times m$ Rayleigh-fading channel, $n << m$, is given by

$$d'_\gamma \approx \frac{(n-r)^2 \ln \frac{\gamma}{a}}{n\left(\frac{1}{m}\|\mathbf{R}_t\|\right)^2}, \quad 0 \leq r \leq n, \qquad (39)$$

where $\mathbf{R}_t$ is the transmit correlation matrix, and $a$ is the SNR offset, all defined in Theorem 2, and $r = nR/\overline{C}$.

*Proof:* see Appendix. ∎

Note the presence of Zheng-Tse term $(n-r)^2$. The $\ln \frac{\gamma}{a}$ term is the average capacity per degree of freedom, which includes the effect of correlation at the Rx end via the SNR offset $a$, and $\frac{1}{m}\|\mathbf{R}_t\|$ is the measure of correlation and power imbalance at the Tx end [26], [29]. Thus, unlike the SNR-asymptotic DMT [6], both of these factors decrease the finite-SNR one. In the absence of correlation (i.e. $\mathbf{R}_r = \mathbf{I}$, $\mathbf{R}_t = \mathbf{I}$), (39) reduces to (37) (with $\beta^* = n/m << 1$, $a \to 1$), as it should be.

### D. Rank-deficient channels

The size-asymptotic approach can also be used for rank-deficient channels via the double-scattering or multi-keyhole models in [26], [29] based on Theorem 3.

*Theorem 6:* Under the conditions of Theorem 3, the size-asymptotic DMT of a rank-deficient correlated channel is given for $0 \leq r \leq M \leq \min(m,n)$ as

$$d'_\gamma \approx \frac{(M-r)^2 \ln \frac{\gamma}{a}}{\sum_{k=1}^M \left(\beta_{tk} m^{-2} \|\mathbf{R}_{tk}\|^2 + \beta_{rk} n^{-2} \|\mathbf{R}_{rk}\|^2\right)}, \quad (40)$$

where $a$ is the SNR offset in Theorem 3 and $r = M \cdot R/\overline{C}$.

*Proof:* see Appendix. ∎

Note that this expression has a structure very similar to that in (39). In particular, the diversity gain decreases with the measures of correlation and power imbalance at both ends $m^{-2}\|\mathbf{R}_{tk}\|^2$, $n^{-2}\|\mathbf{R}_{rk}\|^2$, it is proportional to the Zheng-Tse-like term $(M-r)^2$ and to the mean capacity per degree of freedom $\ln \frac{\gamma}{a}$. Extensive numerical experiments show that (40) provides a reasonable approximation to the true DMT at low to moderate SNR range. Furthermore, unlike the SNR-asymptotic approach, the size-asymptotic one provides a reasonably-accurate estimate of the outage probability at low to moderate SNR range and, thus, can be used as a design criterion for practically-important SNR ranges.

## IV. CONCLUSION

While the SNR-asymptotic DMT is an elegant framework to compare various MIMO systems and channels and also to obtain a number of design guidelines, its use at finite SNR has a number of limitations, which are discussed in this paper. To overcome these limitations, the finite-SNR DMT is obtained for a broad class of fading channels, including full-rank and rank-deficient (double-scattering, keyhole, relay) ones, based on recent results on size-asymptotic outage capacity distribution in such channels. Since the DMT alone is not sufficient to characterize adequately the outage probability, the SNR offset has been introduced and characterized via the size-asymptotic theory. The size-asymptotic, finite-SNR DMT in combination with the SNR offset can be used to characterize accurately the outage probability and also to produce some design guidelines valid at realistic SNR values, including such effects as correlation and power imbalance in the channel. All results and conclusions have been validated via extensive Monte-Carlo simulations. Overall, the size-asymptotic approach is a viable alternative to the SNR-asymptotic one since the former produces the results that hold at realistic SNR and for a broad class of fading distributions (i.e. robust), and include the effect of correlation and power imbalance in the channel.

## APPENDIX

*Proof of Theorem 3:* the following Lemma is instrumental.

*Lemma 1:* Let $\mathbf{h} = \mathbf{R}^{1/2}\mathbf{g}$, where $\mathbf{g}$ is a zero mean, unit variance, i.i.d. complex random vector, and $\mathbf{R}$ is a positive semidefinite (correlation) matrix. Then, the mean and variance of $\|\mathbf{h}\|^2$ are

$$E(\|\mathbf{h}\|^2) = tr\mathbf{R},$$
$$m_2(\|\mathbf{h}\|^2) = \|\mathbf{R}\|^2 m_2|g|^2), \quad (41)$$

where $g$ is any entry of $\mathbf{g}$. If $\mathbf{g}$ is Gaussian, then $m_2(|g|^2) = 1$.

*Proof:*
$$E(\|\mathbf{h}\|^2) = tr\left(\mathbf{R}^{1/2} E(\mathbf{g}\mathbf{g}^+) \mathbf{R}^{1/2}\right) = tr\mathbf{R}$$

$$m_2(\|\mathbf{h}\|^2) = E\left(tr\left(\mathbf{R}^{1/2}(\mathbf{g}\mathbf{g}^+ - \mathbf{I})\mathbf{R}^{1/2}\right)^2\right)$$

$$= E\left(\sum_i \left(|g_i|^2 - 1\right) \lambda_i\right)^2$$

$$= \sum_i E\left(|g_i|^4 - 1\right) \lambda_i^2$$

$$= \|\mathbf{R}\|^2 m_2(|g|^2), \quad (42)$$

where $\lambda_i$ are the eigenvalues of $\mathbf{R}$ and we have used the fact that $E(\mathbf{g}\mathbf{g}^+) = \mathbf{I}$ ∎

Without loss of generality, we assume that $\mathbf{h}_{tk}$ and $\mathbf{h}_{rk}$ are normalized: $m^{-1}tr\{\mathbf{R}_{tk}\} = n^{-1}tr\{\mathbf{R}_{rk}\} = 1$. The channel of Theorem 3 is the multikeyhole model in [[26], Theorem 7], whose instantaneous capacity converges in probability to

$$C \xrightarrow{p} \sum_{k=1}^M \ln\left(1 + \frac{|b_k|^2 \gamma}{m \cdot n} \|\mathbf{h}_{tk}\|^2 \|\mathbf{h}_{rk}\|^2\right), \text{ as } m,n \to \infty, \quad (43)$$

if (i) $\lim_{m \to \infty} m^{-1}\|\mathbf{h}_{tk}\|^2 < \infty$, $\lim_{n \to \infty} n^{-1}\|\mathbf{h}_{rk}\|^2 < \infty$ and (ii) $\lim_{m \to \infty} m^{-2} tr[\mathbf{R}_{tk}\mathbf{R}_{tl}] = 0$, $\lim_{n \to \infty} n^{-2} tr[\mathbf{R}_{rk}\mathbf{R}_{rl}] = 0$ for every $k,l = 1...M$. Condition (i) follows from (b) in Theorem 3. Using Von-Neumann trace [37] and Cauchy-Schwarz inequalities, $tr[\mathbf{R}_1\mathbf{R}_2] \leq \|\mathbf{R}_1\| \cdot \|\mathbf{R}_2\|$, where $\mathbf{R}_1$ and $\mathbf{R}_2$ are positive semidefinite. Thus, a sufficient condition for (ii) to hold is $\lim_{m \to \infty} m^{-1}\|\mathbf{R}_{tk}\| = 0$, $\lim_{n \to \infty} n^{-1}\|\mathbf{R}_{rk}\| = 0$. From [[25], Comment 5], $\left(n^{-1}\|\mathbf{R}\|\right)^{1/3} \leq \|\lambda\|_3 / \|\lambda\|_2$, where $\|\lambda\|_\delta = \left(\sum_i \lambda_i^\delta\right)^{1/\delta}$, and from [[25], Theorem 1], $\|\lambda\|_3 / \|\lambda\|_2 \to 0$ iff $\|\mathbf{R}\|_2 / \|\mathbf{R}\| \to 0$ as $n \to \infty$, so that (ii) holds under condition (c) of Theorem 3.

Let $C_k = \ln(1 + \frac{|b_k|^2 \gamma}{m \cdot n}\|\mathbf{h}_{tk}\|^2 \|\mathbf{h}_{rk}\|^2)$. Under conditions (a) to (c), $C_k$ is asymptotically Gaussian as $m,n \to \infty$ as follows from [[26], Theorem 4], [[25], Theorem 1]. To find the moments of $C_k$, define a function $f(x,y) = \ln(1 + |b_k|^2 \gamma \cdot xy)$ and note that $C_k = f(m^{-1}\|\mathbf{h}_{tk}\|^2, n^{-1}\|\mathbf{h}_{rk}\|^2)$. From the Lemma above and under adopted normalization $m^{-1} tr\{\mathbf{R}_{tk}\} = n^{-1} tr\{\mathbf{R}_{rk}\} = 1$, $E(m^{-1}\|\mathbf{h}_{tk}\|^2) = E(n^{-1}\|\mathbf{h}_{rk}\|^2) = 1$. Since $f$ is a smooth function (first-order derivative is continuous) in the neighborhood of $x = y = 1$, using Cramer Theorem [[39], Theorem 7], the mean and the variance of $C_k$ as $m,n \to \infty$ are

$$\overline{C}_k = f(1,1) = \ln\left(1 + |b_k|^2 \gamma\right) \quad (44)$$

and

$$\sigma_k^2 = \left[\left.\frac{\partial f(x,y)}{\partial x}\right|_{x=1,y=1}\right]^2 \frac{\beta_{tk}}{m^2} \|\mathbf{R}_{tk}\|^2$$
$$+ \left[\left.\frac{\partial f(x,y)}{\partial y}\right|_{x=1,y=1}\right]^2 \frac{\beta_{rk}}{n^2} \|\mathbf{R}_{rk}\|^2$$
$$= \left(\frac{|b_k|^2 \gamma}{1 + |b_k|^2 \gamma}\right)^2 \left(\frac{\beta_{tk}}{m^2} \|\mathbf{R}_{tk}\|^2 + \frac{\beta_{rk}}{n^2} \|\mathbf{R}_{rk}\|^2\right) \quad (45)$$

where $\beta_{tk} = m_2(|g_{tk}|^2)$, $\beta_{rk} = m_2(|g_{rk}|^2)$, and $g_{t(r)k}$ is any entry of $\mathbf{g}_{t(r)k}$. If $\mathbf{g}_{t(r)k}$ are Gaussian, then $\beta_{tk} = \beta_{rk} = 1$. Since $C$ converges to a sum of $C_k$, which are asymptotically Gaussian and independent (due to the mode independence), $C$ is asymptotically Gaussian with the mean and variance

$$\overline{C} = \sum_{k=1}^{M} \overline{C}_k; \ \sigma_C^2 = \sum_{k=1}^{M} \sigma_k^2 \qquad (46)$$

and (18) follows. The approximations follow in a straightforward way. ∎

*Proof of Proposition 1:* From (6), $c(r) = \gamma^{d(r)} P_{out}$. For $0 < \gamma < \infty$, $\lim_{r \to 0} \gamma^{d(r)} < \infty$ and $\lim_{r \to 0} P_{out} = \lim_{\varepsilon \to 0} F_C(\varepsilon) = 0$. Therefore, $c(r) \to 0$ as $r \to 0$. ∎

*Proof of Proposition 2:* We use the following upper-bound,

$$P_{out}(R) = Q\left(\frac{\overline{C} - R}{\sigma_C}\right) \leq \frac{1}{2} \exp\left(-\frac{1}{2}\left(\frac{\overline{C} - R}{\sigma_C}\right)^2\right) \qquad (47)$$

On the *log-log* scale, the gap between $P_{out}$ and the upper-bound is almost a constant, so that the diversity gain is accurately captured by the upper bound. By substituting this upper-bound in (20), (23) and after some manipulations, one obtains (25), (26). ∎

*Proof of Theorem 4:* For square channels, $\beta = 1$, from Theorem 1,

$$\frac{\overline{C}}{n} = 2\ln\left(1 + \gamma - \frac{F(\gamma,1)}{4}\right) - \frac{F(\gamma,1)}{4\gamma},$$
$$\sigma_C^2 = -\ln\left(1 - \left(\frac{F(\gamma,1)}{4\gamma}\right)^2\right), \qquad (48)$$

where $F(\gamma,1) = \left(\sqrt{4\gamma+1} - 1\right)^2$. It is straightforward to see that

$$\frac{F(\gamma,1)}{4\gamma} = 1 - \frac{1}{\sqrt{\gamma}} + \frac{1}{2\gamma} + o\left(\frac{1}{\gamma}\right),$$
$$\frac{F(\gamma,1)}{4} = \gamma - \sqrt{\gamma} + \frac{1}{2} + o(1), \qquad (49)$$

so that, after some manipulations,

$$\frac{\overline{C}}{n} = \ln\left(\frac{\gamma}{e}\right) + \frac{2}{\sqrt{\gamma}} + o\left(\frac{1}{\sqrt{\gamma}}\right),$$
$$\sigma_C^2 = \frac{1}{2}\ln\frac{\gamma}{4} + \frac{1}{\sqrt{\gamma}} + o\left(\frac{1}{\sqrt{\gamma}}\right) \qquad (50)$$

from which (13) follows. From this and using $R = r\overline{C}/n$, one obtains:

$$\left(\frac{\overline{C} - R}{\sigma_C}\right)^2 = (n-r)^2 \frac{\left(\ln\left(\frac{\gamma}{e}\right) + \frac{2}{\sqrt{\gamma}}\right)^2}{\frac{1}{2}\ln\frac{\gamma}{4} + \frac{1}{\sqrt{\gamma}}} + o\left(\frac{1}{\sqrt{\gamma}}\right)$$
$$= 2(n-r)^2 \left(\ln\left(\frac{\gamma}{e}\right) + \frac{2}{\sqrt{\gamma}}\right) + o(1) \qquad (51)$$

Substituting (51) into the upper bound in (47) and after some lengthy by straightforward manipulations, keeping only the lower-order (dominating) terms, one obtains

$$P_{out} \approx \frac{1}{2}\left(\frac{\gamma}{e}\right)^{-d(r)\Delta(\gamma)}, \qquad (52)$$

where $d(r) = (n-r)^2$ and $\Delta(\gamma)$ quantifies the effect of finite SNR,

$$\Delta(\gamma) \approx 1 + 2/\left(\sqrt{\gamma}\ln(\gamma/e)\right) \qquad (53)$$

Interpreting the $1/e$ term in (52) as a high-SNR offset (similarly to [33]), the diversity gain in (20) becomes $d_\gamma \approx d(r)\Delta(\gamma)$. Using (52), the differential diversity gain (23) can be expressed as $d'_\gamma = d(r)(\Delta(\gamma) + \gamma\ln(\gamma/e)\partial\Delta(\gamma)/\partial\gamma)$, which, after some manipulations, can be simplified to (27).

While the upper bound in (47) is of sufficient accuracy to evaluate the diversity gains, a more refined approximation is required to capture accurately the SNR offset,

$$Q(z) = \frac{1}{\sqrt{2\pi}z}\exp\left(-\frac{z^2}{2}\right) + o\left(\frac{1}{z}\exp\left(-\frac{z^2}{2}\right)\right) \qquad (54)$$

from which the outage probability can be approximated, for $d_\gamma \ln(\gamma/e) > 1$ and $r > 0$, as

$$P_{out} \approx \frac{1}{\sqrt{4\pi d_\gamma \ln(\gamma/e)}}\left(\frac{\gamma}{e}\right)^{-d_\gamma} \qquad (55)$$

Using this in (24), the SNR offset becomes

$$c_\gamma \approx \frac{e^{d_\gamma}e^{(d'_\gamma - d_\gamma)\ln\gamma}}{\sqrt{4\pi d_\gamma \ln(\gamma/e)}} \qquad (56)$$

It is straightforward to see that $d'_\gamma - d_\gamma = o(1/\ln\gamma)$, so that (56) simplifies to (28), and (29) follows. Note that the SNR offset in (28) can also be identified by inspection of (55). From Proposition 1, $c_\gamma \to 0$ when $r \to 0$, and $d_\gamma = 0$, $c_\gamma = 1/2$ for $r = n$ by inspection of (19). ∎

*Proof of Proposition 4 (sketch):* This follows mostly the steps of that of Theorem 4: Using (13) and (21), the outage probability can be approximated, via the upper bound in (47), as

$$P_{out} \approx \frac{1}{2}\left(\frac{\gamma}{a}\right)^{-d_\gamma}, \qquad (57)$$

where

$$d_\gamma \approx \frac{1}{2}(m^* - r)^2 \frac{\ln(\gamma/a)}{-\ln(1-\beta^*)} \qquad (58)$$

is the diversity gain. Substituting (57) into (23) gives $d'_\gamma$ in (37). ∎

*Proof of Theorem 5 (sketch):* Following Theorem 2, we use the approximation $\sigma_C^2 \approx \frac{n}{m^2}\|\mathbf{R}_t\|^2$, assuming $rank(\mathbf{R}_r) = n$, to evaluate the capacity variance. Using the multiplexing gain definition in (21), the outage probability at moderate to high SNR can be approximated via the upper bound in (47) as in (57), where

$$d_\gamma \approx \frac{(n-r)^2}{2n}\frac{\ln\frac{\gamma}{a}}{\left(\frac{1}{m}\|\mathbf{R}_t\|\right)^2}, \qquad (59)$$

is the finite-SNR diversity gain. Using this in (23), (31) follows after some straightforward manipulations. ∎

*Proof of Theorem 6 (sketch):* Using the high SNR approximation in (18), the outage probability can be approximated as in (57) with

$$d_\gamma \approx \frac{(M-r)^2 \ln\frac{\gamma}{a}}{2\sum_{k=1}^{M}\left(\frac{\beta_{tk}}{m^2}\|\mathbf{R}_{tk}\|^2 + \frac{\beta_{rk}}{n^2}\|\mathbf{R}_{rk}\|^2\right)}, \qquad (60)$$

where the SNR offset $a$ is as in Theorem 3. Using this in (23), (40) follows. ∎

**Sergey Loyka** (M'96–SM'04) was born in Minsk, Belarus. He received the Ph.D. degree in Radio Engineering from the Belorussian State University of Informatics and Radioelectronics (BSUIR), Minsk, Belarus in 1995 and the M.S. degree with honors from Minsk Radioengineering Institute, Minsk, Belarus in 1992. Since 2001 he has been a faculty member at the School of Information Technology and Engineering, University of Ottawa, Canada. Prior to that, he was a research fellow in the Laboratory of Communications and Integrated Microelectronics (LACIME) of Ecole de Technologie Superieure, Montreal, Canada; a senior scientist at the Electromagnetic Compatibility Laboratory of BSUIR, Belarus; an invited scientist at the Laboratory of Electromagnetism and Acoustic (LEMA), Swiss Federal Institute of Technology, Lausanne, Switzerland. His research areas include wireless communications and networks, MIMO systems and smart antennas, RF system modeling and simulation, and electromagnetic compatibility, in which he has published extensively. Dr. Loyka is a technical program committee member of several IEEE conferences and a reviewer for numerous IEEE periodicals and conferences. He received a number of awards from the URSI, the IEEE, the Swiss, Belarus and former USSR governments, and the Soros Foundation.

**Georgy Levin** received the B.S. and M.S. degrees, both cum laude, in Electrical and Computer Engineering from Ben-Gurion University of the Negev, Israel in 1995 and 2000, and the Ph.D. degree from the University of Ottawa, Ontario, Canada in 2008. He is currently a research assistant at the University of Ottawa. Dr. Levin's research spans the fields of wireless communications and information theory with specific interest in MIMO systems, smart antennas, relay networks, cognitive radio and synthetic aperture radars.